\documentclass{moriond}

\def\be{\begin{equation}}
\def\ee{\end{equation}}
\def\bea{\begin{eqnarray}}
\def\eea{\end{eqnarray}}

\def\MET{E$_{\rm T}^{\rm miss}\,$}

\newcommand{\Photo}{}

\begin{document}
\vspace*{4cm}
\title{Searches for weakly produced SUSY at LHC}

\author{ S. Folgueras for the CMS and ATLAS Collaborations }

\address{Departmento de F\'isica, Universidad de Oviedo, \\
C/Calvo Sotelo s/n, 33005, Oviedo, Spain}

\maketitle\abstracts{
A summary of the different searches for weakly produced SUSY by both CMS and ATLAS is presented here. A review on the methodology of these searches, including event selection, background suppression and estimation methods, etc is covered. Other searches at the LHC already probe squarks and gluino masses up to 1.4 TeV, such scenario, may favour electroweak production of charginos and neutralinos, that will produce many-lepton final states accompanied by E$_{\rm T}^{\rm miss}\,$ and very little hadronic activity. Latest searches include Higgs boson in the decay and exploits VBF associated production to probe scenarios with very small mass splittings.
}

\section{Introduction}
SUSY models are a possible extension to the very successful standard model (SM) that provides a solution to many problems such as fine-tuning and the unification of the couplings. The search for SUSY has become a major goal and one of the highest priorities searches for both ATLAS~\cite{ATLAS} and CMS~\cite{CMS}.

While most of LHC searches focus on strong production of SUSY with larger cross sections, electroweak production of SUSY may be the key to new physics, as the squarks and gluinos may be too heavy to be produced at LHC energies. The decay chain will produce sleptons (or W,Z and h bosons) and lead to multiple-lepton final states with little hadronic activity and significant missing transverse energy (\MET) from the lightest-supersymetric particle (LSP).

As the production cross-section is very small, many final states are combined together targeting different production mechanisms to reach enough sensitivity for new physics. A summary of the different searches performed by CMS and ATLAS Collaborations will be shown in these proceedings.  

\section{Signature}
Very clean lepton signatures with low hadronic activity are expected in the final state when assuming direct chargino-neutralino production with both light and heavy sleptons. In the second case, the chargino/neutralino decays directly to W,Z or h bosons. Three lepton final states~\cite{CMS_EWKino,ATLAS_3L} are yielded naturally by most of these models, however, under certain conditions, two lepton signatures could be more sensitive, i.e. when the mass splitting is small. 
On top of the multilepton selection, a Z-veto is generally applied to suppress background from diboson production, a (b)jet-veto is applied to suppress $\rm t \bar{t}$ background. Signal regions are defined in bins of \MET, $\rm M_{T(2)}$ and $\rm m_{ll}$. Two remaining backgrounds have to be estimated: non-prompt lepton backgrounds are estimated using different data-driven techniques. Irreducible diboson backgrounds are estimated using simulation. Control regions are used for MC validation.

\section{Chargino-neutralino production}
Previous searches from both CMS and ATLAS have looked for chargino-neutralino production decaying via sleptons or W,Z bosons (if the sleptons are too heavy). Many final states were used to probe electroweak production of charginos and neutralinos: the three leptons~\cite{CMS_EWKino,ATLAS_3L} final state is the most natural signature. If the mass splitting between the particles is small, the sensitivity of the three-lepton analysis is very poor, however some of this efficiency can be recovered by dilepton final sates (both opposite-sign~\cite{ATLAS_2L} and same-sign~\cite{CMS_EWKino}). Finally, depending on the considered model, a final state with multiple taus can be hugely favoured, therefore a dedicated tau search~\cite{ATLAS_2tau} is also particularly sensitive to these scenarios.

The Higgs discovery opened up a new possibility in electroweak production of SUSY, the neutralino could decay, not only into sleptons, or W/Z bosons, but also into a Higgs boson, leading to many new different final states that could be used to probe for electroweak production, both considering a light or a heavy LSP. 

Considering different Higgs-decay-modes, various searches targeting chargino-neutralino production will be summarized. A W boson is produced in the chargino decay chain, while the neutralino decays to a Higgs boson and the LSP. 

\begin{figure}[ht]
  \begin{minipage}{0.32\textwidth}
    \centerline{\includegraphics[width=\linewidth,height=4cm]{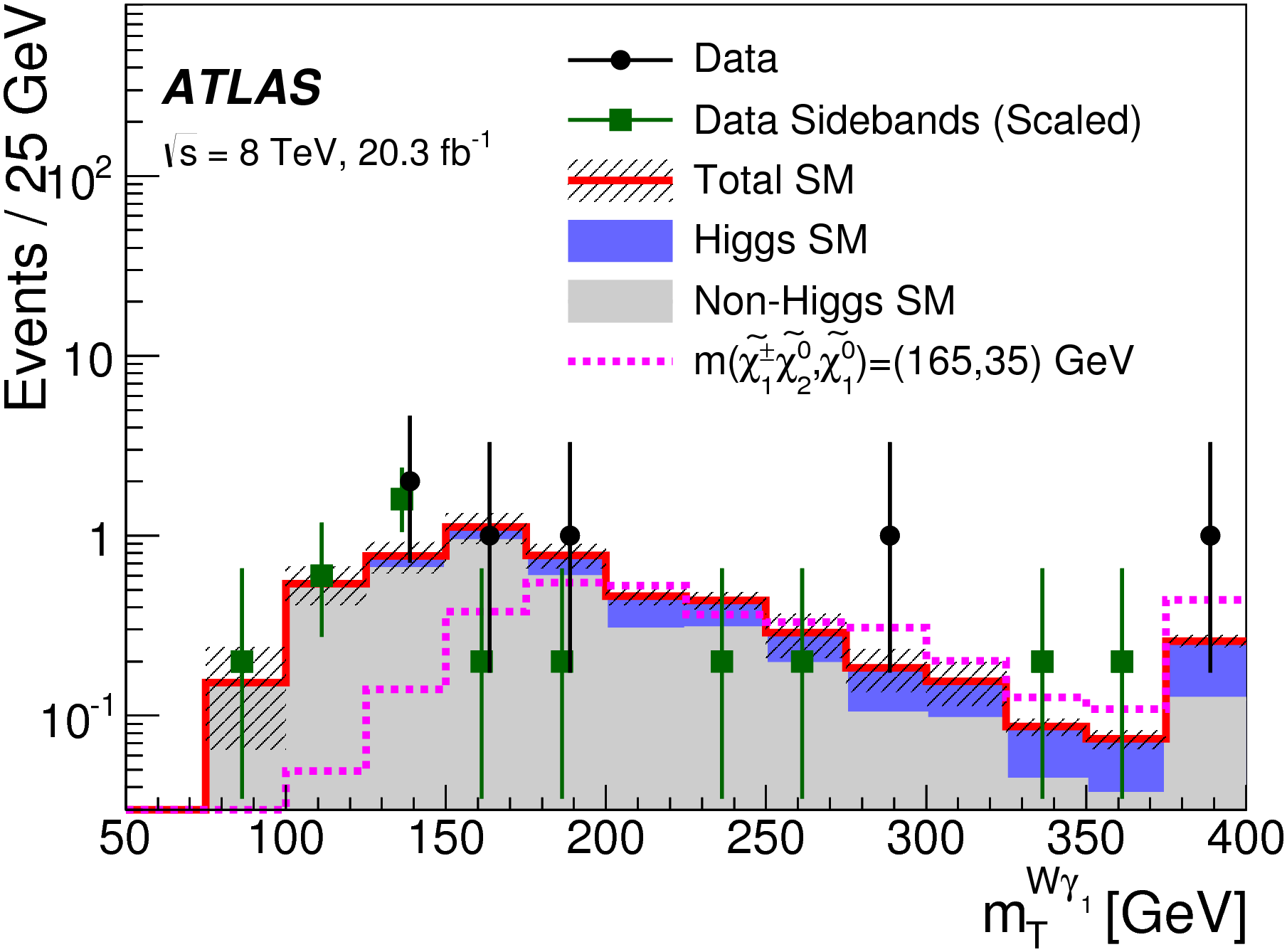}}
  \end{minipage} 
  \hfill
  \begin{minipage}{0.32\textwidth}
    \centerline{\includegraphics[width=\linewidth,height=4cm]{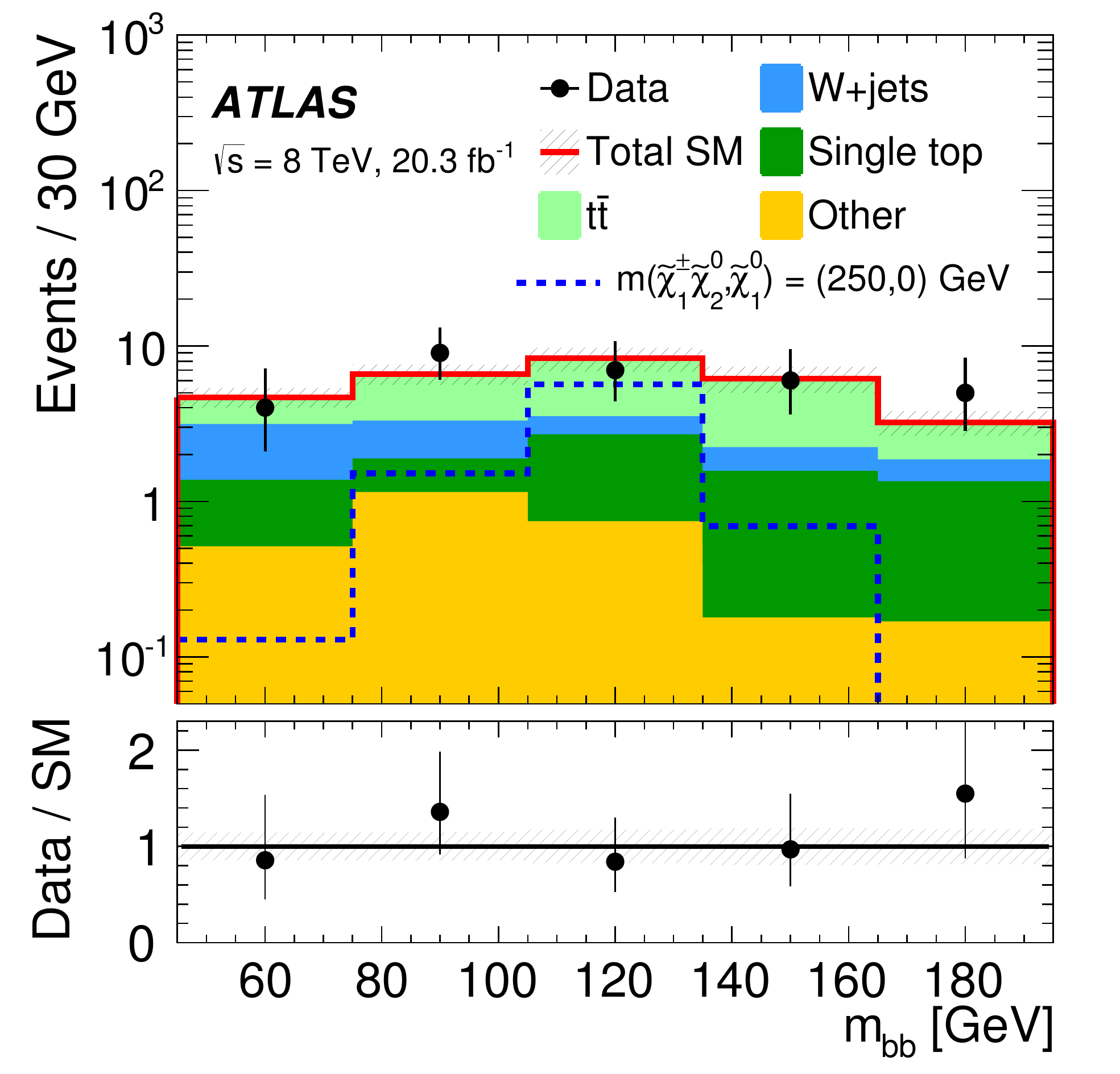}}
  \end{minipage}
  \hfill
  \begin{minipage}{0.32\textwidth}
    \centerline{\includegraphics[width=\linewidth,height=4cm]{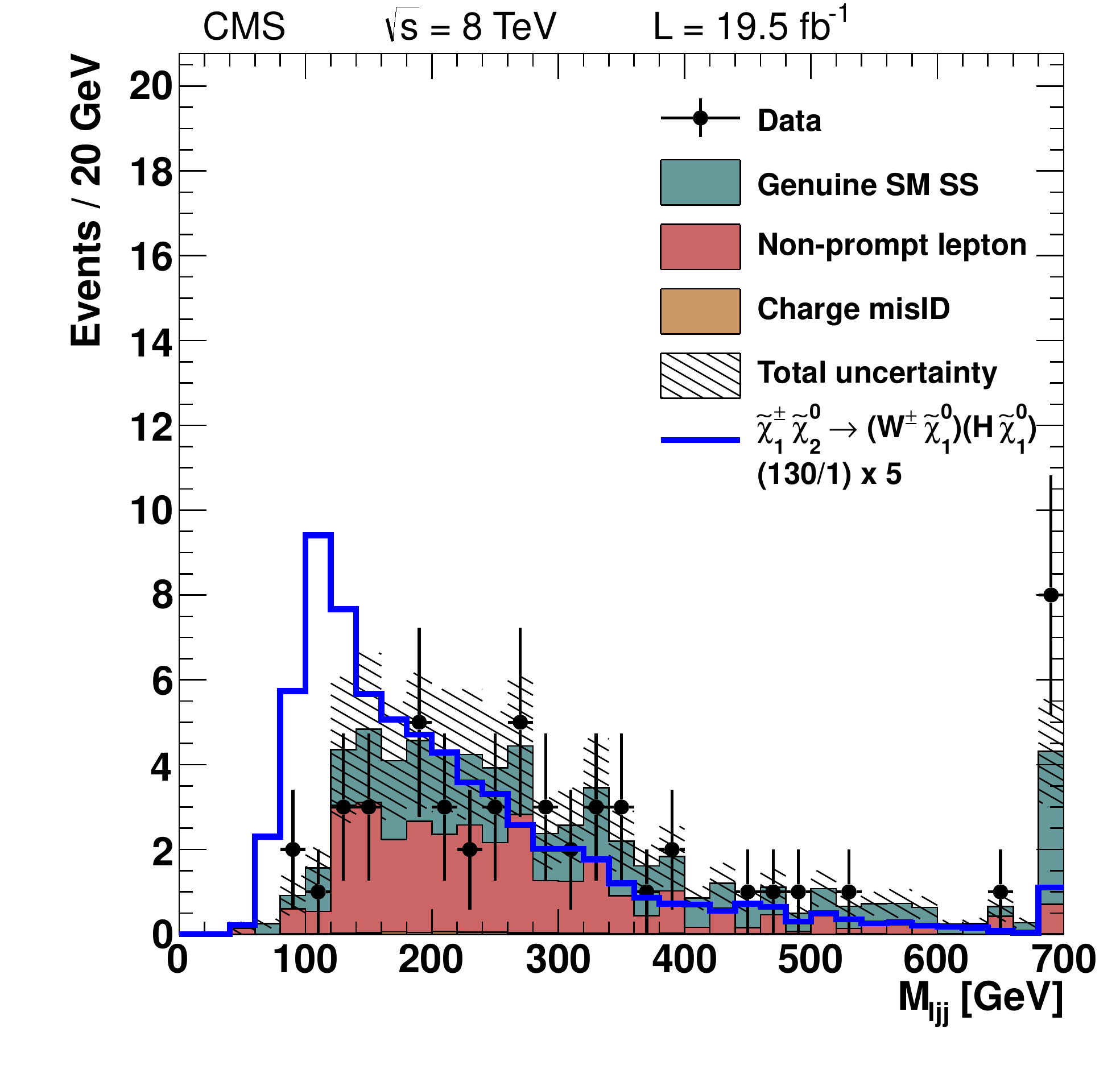}}
  \end{minipage}
  \caption[]{Transverse mass distribution in the $\gamma\gamma+$lepton search~\cite{ATLAS_H} (left), $m_{bb}$ distribution in the single-lepton search~\cite{ATLAS_H} (middle) and $m_{ljj}$ for the same-sign search~\cite{CMS_EWKino} (right). All three searches target $\tilde{\chi}^\pm_1\tilde{\chi}^0_2 \rightarrow\ hW^\pm\ \tilde{\chi}^0_1 \tilde{\chi}^0_1 $.}
  \label{fig:tchi}
\end{figure}

\subsection{\texorpdfstring{$\gamma\gamma$} + lepton search}
When the Higgs decays to two photons, the final state consists of two photons and a lepton. The signal region is defined using the transverse mass as a discriminating variable. The transverse mass is calculated with respect to the lepton in CMS \cite{CMS_H} case and with respect one photon for ATLAS \cite{ATLAS_H} case.  

Non-SM Higgs backgrounds are estimated using a fit on the sidebands of the diphoton invariant mass distribution. 

\subsection{Single lepton search}
To target the $h \rightarrow b \bar{b}$ final state, events with a lepton (coming from the W) and two b-jets from the Higgs decay are selected. The search strategy divides the signal region into \MET \cite{CMS_EWKino} and $\rm m_{CT}, m_T$ bins \cite{ATLAS_H} and looks for a resonance in the $m_{bb}$ spectrum.  

Background is estimated from MC, however, several control regions are used to constrain normalization and validation regions are used test background modelling. 

\subsection{Same-sign search}
For searching in the $h \rightarrow W⁺W⁻$ channel, events with two same-sign leptons, no b-jets and 1,2 or 3 jets are selected. Extra cuts on \MET and $\rm m_{T2}$ are then applied to remove background contribution in the CMS search \cite{CMS_EWKino} while ATLAS \cite{ATLAS_H} uses extra cuts on \MET, $|\Delta\eta|$, $m_T$ and $m_{eff}$. 

The search strategy is very similar, looking for a peak in the low $m_{ljj}$ region. 

\subsection{Combination}
As no significant excess over the Standard Model predictions is seen in either channel, a combination of all these searches together with the previous multilepton result\cite{ATLAS_3L,ATLAS_2tau,CMS_EWKino} to set limits on chargino-neutralino production when $\tilde{\chi}^0_2 \rightarrow h \tilde{\chi}^0_1$. Chargino masses up to 250 (210) GeV are probed in ATLAS(CMS) search. 

\begin{figure}[ht]
  \begin{minipage}{0.49\textwidth}
    \centerline{\includegraphics[height=3.7cm]{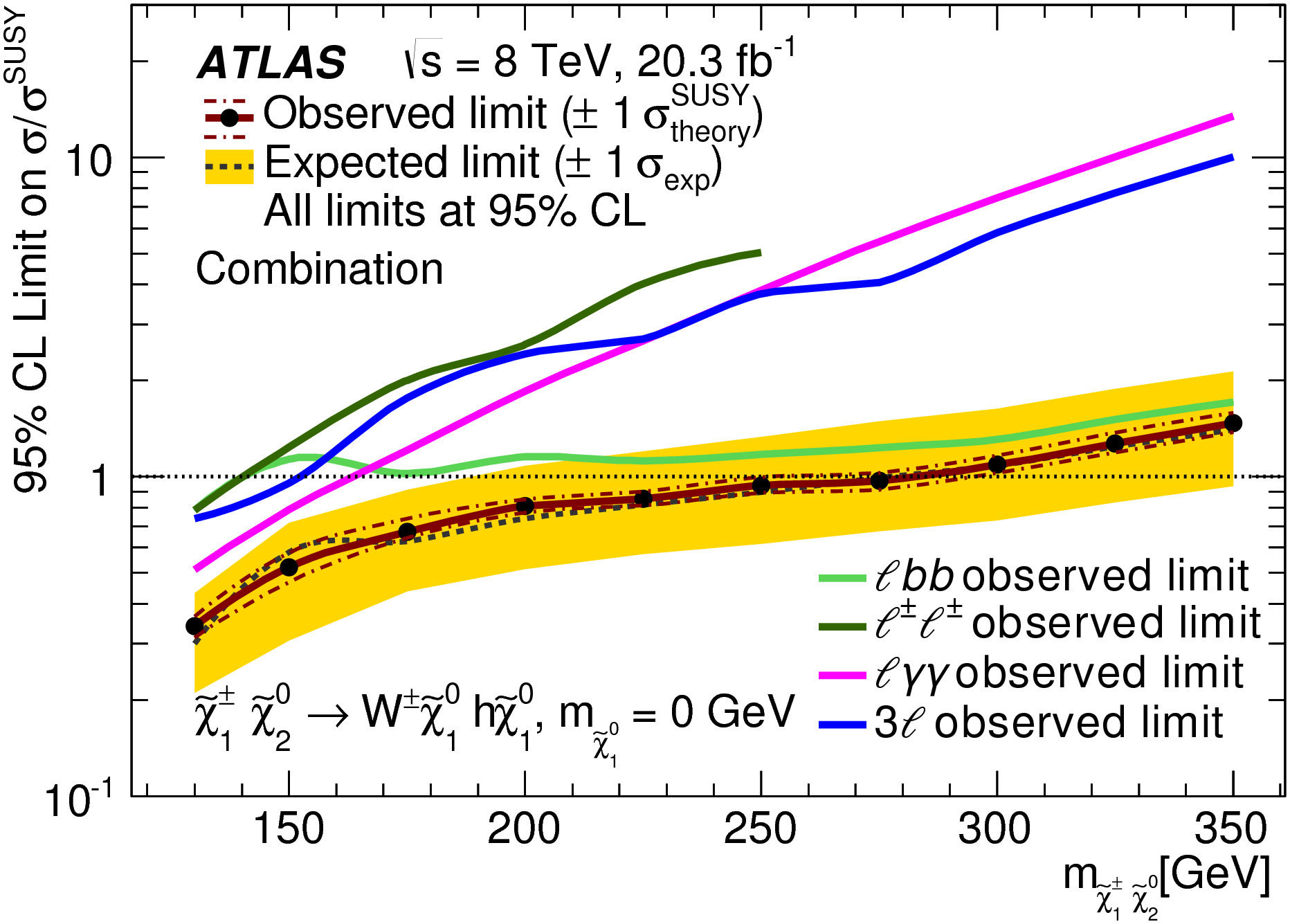}}
  \end{minipage} 
  \begin{minipage}{0.49\textwidth}
    \centerline{\includegraphics[height=4.1cm]{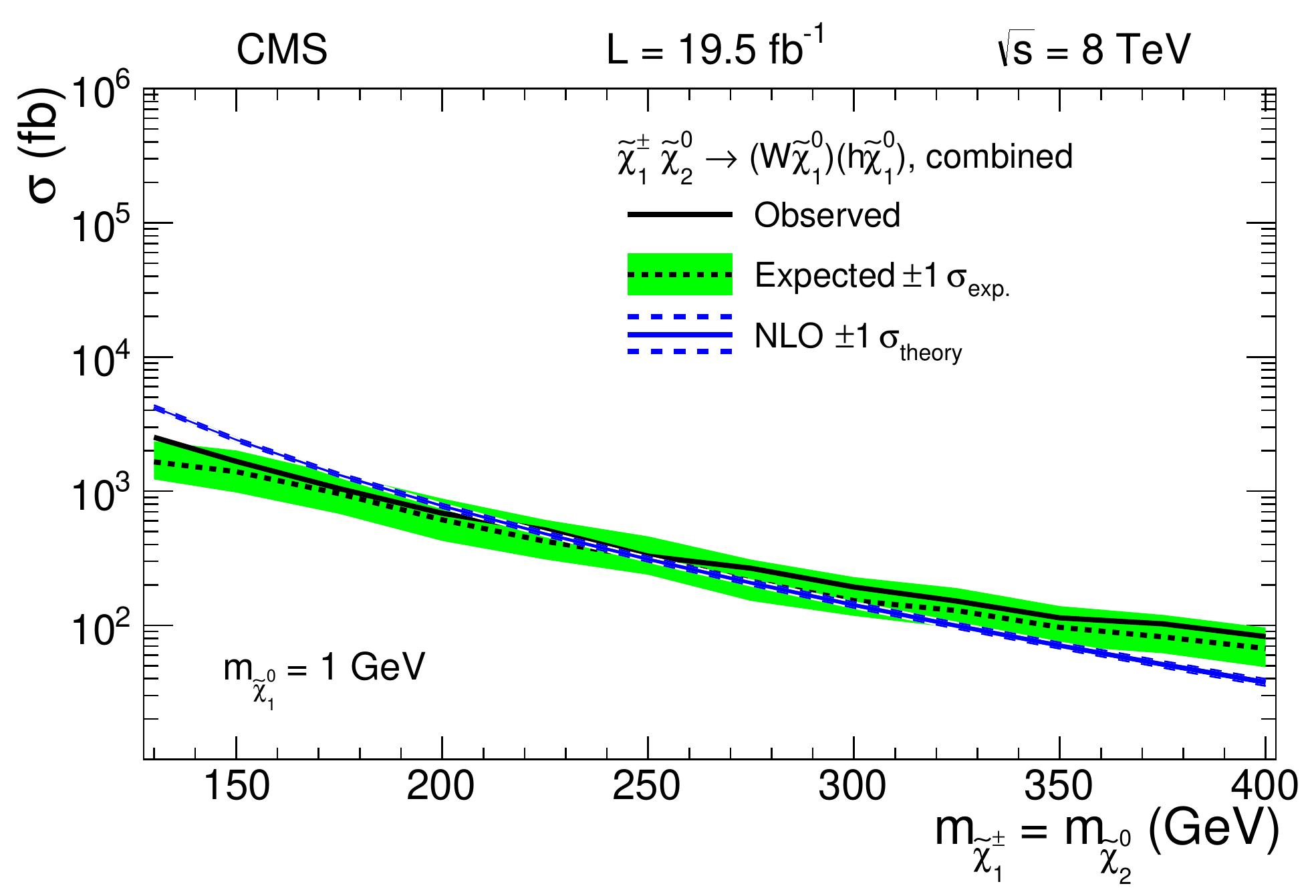}}
  \end{minipage}
  \caption[]{Left: Observed (solid line) and expected (dashed line) 95\% CL upper limits on the cross section normalised by the simplified model prediction as a function of the common mass $m_{\tilde{\chi}^{\pm}_1\tilde{\chi}^0_2}$ for  $\tilde{\chi}^0_1=0$ GeV for the ATLAS combination~\cite{ATLAS_H}. Right: Observed and expected 95\% CL upper limits on the cross section as a function of the mass $m_{\tilde{\chi}^{\pm}_1\tilde{\chi}^0_2}$ for $\tilde{\chi}^0_1=1$ GeV for CMS combination~\cite{CMS_H}. In both cases, the solid band around the expected limit represents the one-standard-deviation interval.}
  \label{fig:tchi_summary}
\end{figure}

\section{Electroweak production of SUSY with VBF production}
More recent searches exploit the VBF topology, allowing a complementary route to probe the electroweak production sector, specially when trying to reach very compressed scenarios that are not accessible with more classic searches. 

\subsection{Photon + \MET search}
Events with one or two photons, no lepton, two VBF jets (large $|\Delta\eta|$ gap between the two jets) and \MET are selected in this analysis~\cite{ATLAS_VBF}. Signal regions are designed on top of this selection to fully exploit the VBF topology, the invariant mass  $|\Delta\eta|$ of the dijet system is used for the search region definition.

Background processes are mainly multijets and $\gamma$+jets, they are estimated directly from data control regions. Having found no significant excess over the SM expectations, limits are set for a various of LSP and NLSP masses in a GSMB model where the Higgs boson decays to a pair of neutralinos. Although the analysis was optimized for the one photon+\MET signature, most stringent limits arise from the di-photon search. 

\subsection{Two lepton search}
In this CMS analysis \cite{CMS_VBF}, events with 2 leptons, \MET and two VBF jets (large $|\Delta\eta|$ gap between the two jets) are selected. A broad excess in the tails of the invariant mass distribution of the two jets will be an indication for new physics.  

Main background for this search arises from $\rm t\bar{t}$, $\rm V+jets$, diboson and QCD, although the tight VBF requirements on the two jets reject most of the backgrounds produced via strong interaction. Simulation is used for estimating the remaining background processes and control-regions in data are used to constrain normalization of each process. 

No excess over the SM prediction is seen in data and limits on chargino-neutralino production both decaying to staus, for different mass splitting and LSP masses are set. Chargino masses up to 250 GeV are excluded for a model with chargino-neutralino production, decaying to staus for a given mass splitting of ($m_{\tilde{\tau}}-m_{\tilde{\chi}^\pm_1}=5$ GeV) and a massless LSP.

\section{Summary}
\begin{figure}[ht]
  \begin{minipage}{0.5\textwidth}
    \centerline{\includegraphics[height=4cm]{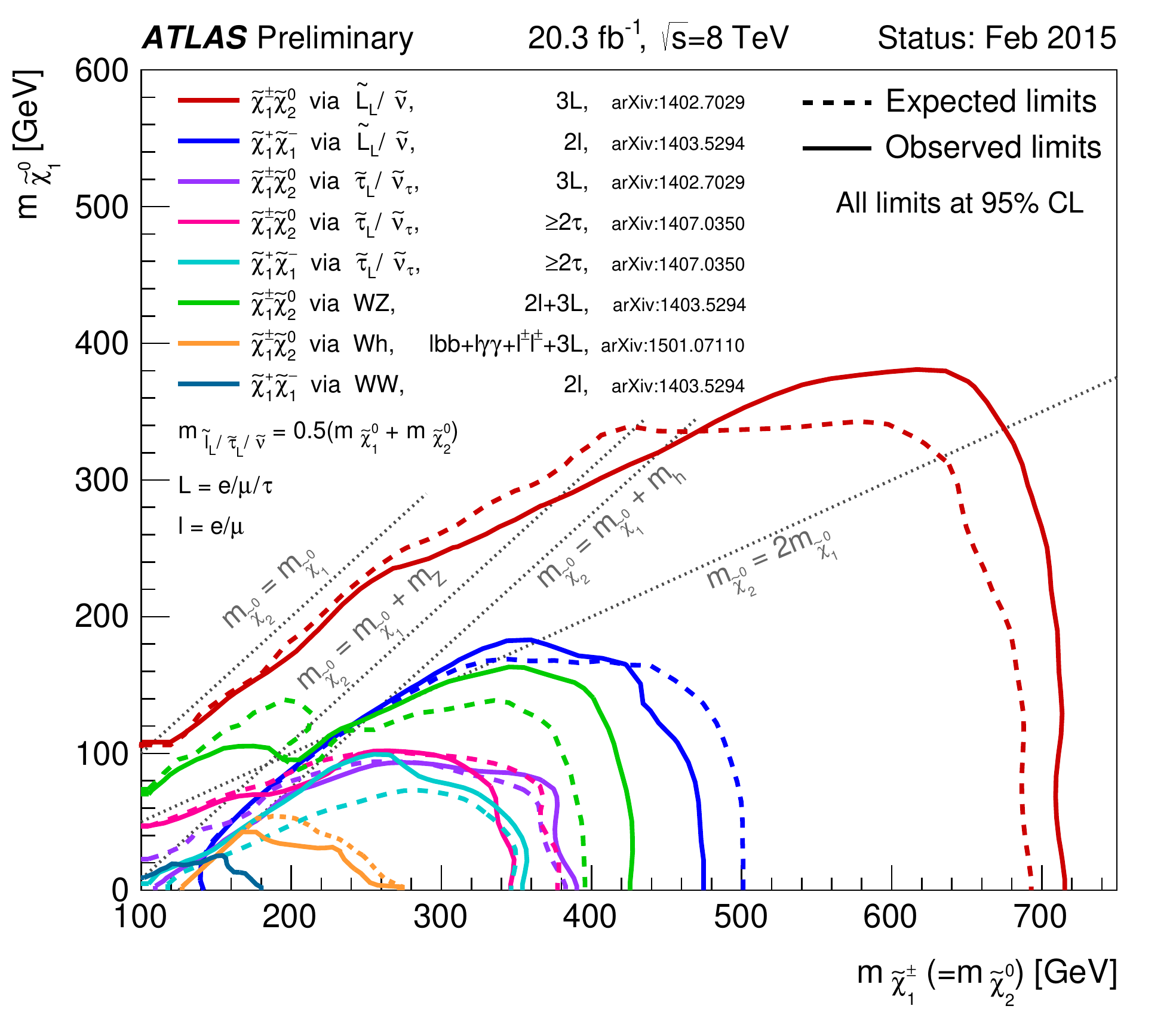}}
  \end{minipage} 
  \begin{minipage}{0.5\textwidth}
    \centerline{\includegraphics[height=4.4cm]{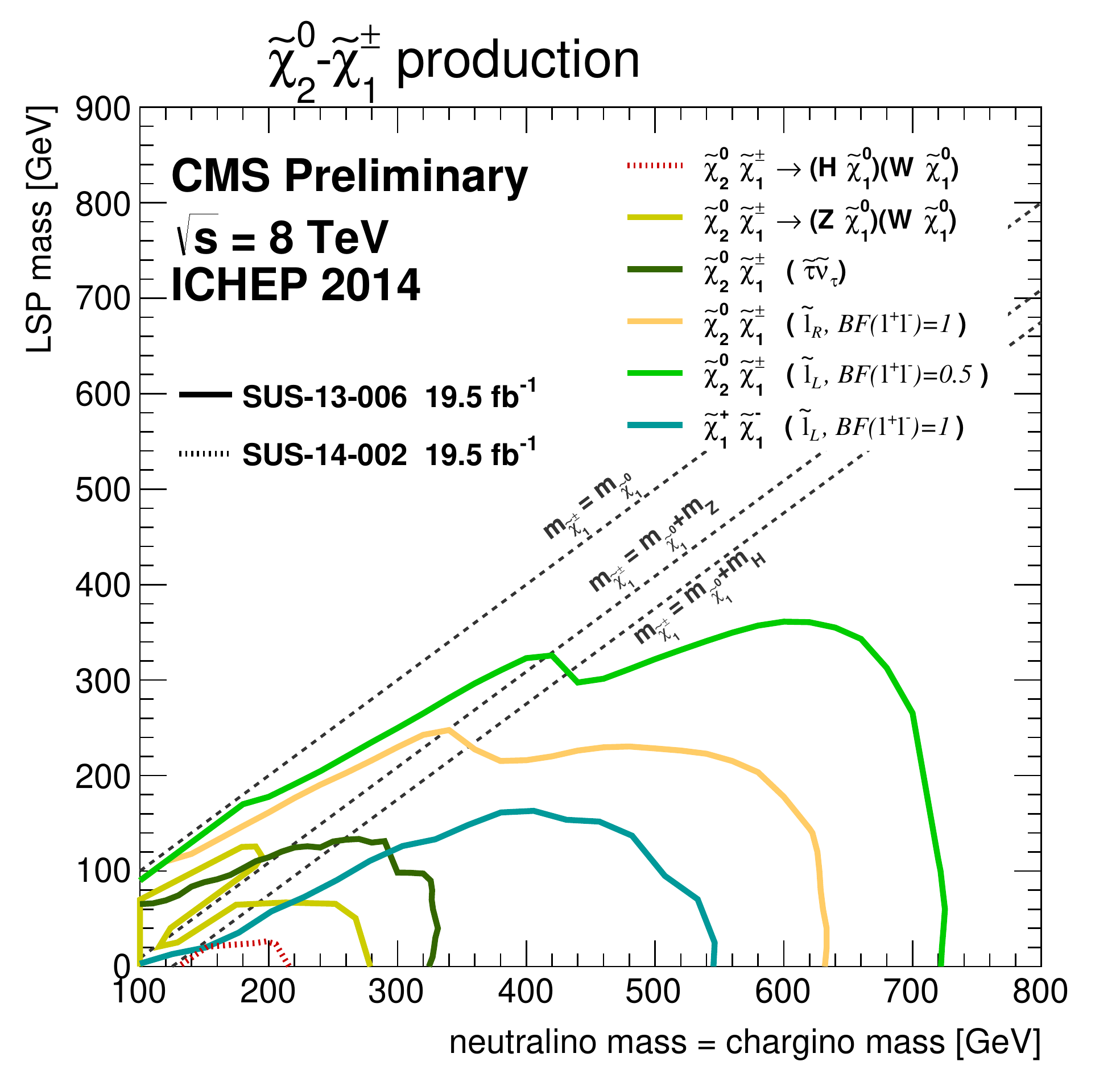}}
  \end{minipage}
  \caption[]{Summary of ATLAS~\cite{ATLAS_public} (left) and CMS~\cite{CMS_public} (right) searches of electroweak production of charginos and neutralinos based of 20 $\rm fb^{-1}$ of pp collision data at $\sqrt{s} = 8$ TeV. Exclusion limits at 90\% confidence level are shown in the chargino-neutralino mass plane for several decay modes of the charginos and neutralinos.}
  \label{fig:Summary}
\end{figure}

A wide variety of searches for weakly produced SUSY have been shown. Current CMS and ATLAS searches probe chargino and neutralino masses up to 700 GeV. Latest searches include Higgs bosons in the decays, probing chargino masses up to 250 GeV. More recent searches also look for weakly produced SUSY in association with two VBF jets, allowing to probe much compressed scenarios up to chargino masses of 250 GeV. The higher energy at the LHC will open up a whole new regime to look for SUSY. Figure~\ref{fig:Summary} shows a the summary of all the searches performed by both collaborations.

\section*{References}
\bibliography{references}

\end{document}